\documentstyle[aps,preprint]{revtex}
\tightenlines
\begin{document}
\draft{}
\title{ Phys. Rev. D 61, 117504 (2000)\\[2cm]
 Final-state interaction phase difference in  $J/\psi\to\rho\eta$ and
$\omega\eta$ decays}
\author{N.N. Achasov
\thanks{Electronic address: achasov@math.nsc.ru}
and V.V. Gubin
\thanks{Electronic address: gubin@math.nsc.ru}
\\
Laboratory of Theoretical Physics,\\
Sobolev Institute for Mathematics,\\
Academician Koptiug prospekt, 4,\\
630090, Novosibirsk 90, Russia}
\date{\today}
\maketitle
\begin{abstract}
It is shown that the study of the $\omega-\rho^0$ interference pattern in
the $J/\psi\to (\rho^0+\omega )\eta\to\pi^+\pi^-\eta$ decay provides
evidence for the large (nearly $90^\circ$) relative phase between the
one-photon  and the three-gluon decay amplitudes.
\end{abstract}


In the last few years it  has been noted that the single-photon and
three-gluon amplitudes in the two-body $J/\psi\to 1^-0^-$ and
$J/\psi\to 0^-0^-$ \cite{castro-95,suzuki-98,suzuki-99}  decays appear to
have relative phases nearly $90^\circ$.
This unexpected result is very important to the observability of CP
violating decays  as well as to the nature of the $J/\psi\to 1^-0^-$ and
$J/\psi\to 0^-0^-$ decays
\cite{castro-95,suzuki-98,suzuki-99,rosner-99,tuan-99,achasov-98,gerard-99}.
Since the analysis \cite{castro-95,suzuki-98,suzuki-99} involved theoretical
assumptions relying on the $SU_f(3)$ symmetry, the strong $SU_f(3)$-symmetry
breaking and so on, the measurements of these phases are urgent.
Fortunately, it is possible to check the conclusion of Refs.
\cite{castro-95,suzuki-98} at least in one case.
We mean the phases between the amplitudes of the one-photon
$J/\psi\to\rho^0\eta$ and three-gluon $J/\psi\to\omega\eta$ decays.

Indeed, the $\omega-\rho$ interference pattern in
$J/\psi\to (\rho^0+\omega )\eta\to\rho^0\eta\to\pi^+\pi^-\eta$ is conditioned
by the $\rho^0-\omega$ mixing and the ratios of the amplitudes of the
$\rho^0$ and $\omega$ production. As for the $\rho^0-\omega$ mixing amplitude,
it is reasonably well studied
\cite{goldhaber-69,gourdin-69,renard-70,achasov-78,achasov-92,achasov-74,pdg-98}. Its
modulus and phase are known. The modules of the ratios of the amplitudes of
the $\rho$ and $\omega$ production can be obtained from the data on the
branching ratios of the $J/\psi$-decays. So, the interference pattern provides
a way of measuring the relative phases of the $\rho^0$ and $\omega$
production amplitudes.

The $\pi^+\pi^-$-spectrum in the $\omega$, $\rho$ energy region is of the
form
\begin{eqnarray}
\label{spectrum}
&& \frac{dN}{dm}= N_\rho (m)\frac{2}{\pi}m\Gamma (\rho\to\pi\pi\,,\, m)\left
|\frac{1}{D_\rho (m)} + \frac{\Pi_{\omega\rho^0}(m)}{D_\rho (m)D_\omega (m)}
\left [\frac{N_\omega (m)}{N_\rho (m)}\right ]^{\frac{1}{2}}
\exp\left\{i\left (\delta_\omega -\delta_\rho\right )\right\} + \right.
\nonumber \\[1pc]
&& \left. + \frac{1}{D_\omega (m)}\cdot\frac{g_{\omega\pi\pi}}
{g_{\rho\pi\pi}}\left [\frac{N_\omega (m)}{N_\rho (m)}\right ]^{\frac{1}{2}}
\exp\left\{i\left (\delta_\omega - \delta_\rho\right )\right\}\right |^2\,,
\end{eqnarray}
where m is the invariant mass of the $\pi^+\pi^-$-state, $N_\rho (m)$ and
$N_\omega (m)$ are the squares of the modules of the $\rho$ and $\omega$
production amplitudes, $\delta_\rho$ and $ \delta_\omega$ are their phases,
$\Pi_{\omega\rho^0}(m)$ is the amplitude of the $\rho-\omega$ transition,
$D_V(m)=m_V^2 - m^2 - im\Gamma_V(m)$, $V=\rho$, $\omega$.

In the discussion that follows, Eq. (\ref{spectrum}) is conveniently
rewritten as
\begin{eqnarray}
\label{spectrum1}
&& \frac{dN}{dm}= N_\rho (m)\frac{2}{\pi}m\Gamma (\rho\to\pi\pi\,,\, m)\left
|\frac{1}{D_\rho (m)}\left (1-\varepsilon (m)\left [\frac{N_\omega (m)}
{N_\rho (m)}\right ]^{\frac{1}{2}}\exp\left\{i\left (\delta_\omega -
 \delta_\rho\right )\right\}\right ) +\right. \nonumber\\[1pc]
&& \left. + \frac{1}{D_\omega (m)}\left (\varepsilon (m)+
g_{\omega\pi\pi}/g_{\rho\pi\pi}\right )\left
[\frac{N_\omega (m)}{N_\rho (m)}\right ]^{\frac{1}{2}}
\exp\left\{i\left (\delta_\omega - \delta_\rho\right )\right\}\right |^2\,,
\end{eqnarray}
where
\begin{eqnarray}
\label{epsilon}
&& \varepsilon (m)=-\frac{\Pi_{\omega\rho^0}(m)}{m_\omega^2-m_\rho^2 +
im\left (\Gamma_\rho (m)-\Gamma_\omega (m)\right )}.
\end{eqnarray}

As known \cite{goldhaber-69,gourdin-69,renard-70,achasov-78,achasov-92},
the imaginary part of the $\rho-\omega$ transition amplitude is due to the
$\pi\pi$, 3$\pi$, $\gamma\pi$ and $\gamma\eta$ intermediate states
\begin{eqnarray}
\label{im}
&& Im\left (\Pi_{\rho^0\omega}(m)\right )= m\left (\frac{g_{\omega\pi\pi}}
{g_{\rho\pi\pi}}\Gamma (\rho\to\pi\pi\,,\, m) + \frac{g_{\rho\rho\pi}}
{g_{\omega\rho\pi}}\Gamma (\omega\to\rho\pi\to 3\pi\,,\, m) +\right.
\nonumber \\
&& \left. + \frac{g_{\rho\gamma\pi}}{g_{\omega\gamma\pi}}\Gamma
(\omega\to\gamma\pi\,,\, m_\omega ) +
\frac{g_{\rho\gamma\eta}}{g_{\omega\gamma\eta}}\Gamma
(\omega\to\gamma\eta\,,\, m ) \right)\,.
\end{eqnarray}
The quite conservative estimate of the contribution of the $\pi\pi$ and
3$\pi$ intermediate states gives
\begin{eqnarray}
\label{pipi3pi}
&& m_\omega \frac{g_{\omega\pi\pi}}{g_{\rho\pi\pi}}
\Gamma (\rho\to\pi\pi\,,\, m_\omega )=\pm m_\omega\cdot 10^{-2}\cdot\Gamma
(\rho\to\pi\pi\,,\, m_\omega )=\pm 1.17\cdot 10^{-3}\,\,\mbox{GeV}^{-2}\,,
\nonumber\\
&& m_\omega  \frac{g_{\rho\rho\pi}}{g_{\omega\rho\pi}}\Gamma
(\omega\to\rho\pi\to 3\pi\,,\, m_\omega )=\pm m_\omega\cdot 10^{-2}\cdot
\Gamma (\omega\to 3\pi\,,\, m_\omega )=\pm 5.84\cdot 10^{-5}\,
\mbox{GeV}^{-2}.
\end{eqnarray}

The constituent quark and vector meson dominance models  both
give the same result
\begin{eqnarray}
\label{gammapigammaeta}
&& m_\omega \frac{g_{\rho\gamma\pi}}{g_{\omega\gamma\pi}}\Gamma
(\omega\to\gamma\pi\,,\, m_\omega)=m_\omega\cdot\frac{1}{3}\cdot\Gamma
(\omega\to\gamma\pi\,,\, m_\omega)=1.86\cdot 10^{-4}\,\,\mbox{GeV}^{-2}\,,
\nonumber\\
&& m_\omega \frac{g_{\rho\gamma\eta}}{g_{\omega\gamma\eta}}\Gamma
(\omega\to\gamma\eta\,,\, m_\omega)=m_\omega\cdot 3\cdot\Gamma
(\omega\to\gamma\eta\,,\, m_\omega)=1.28\cdot 10^{-5}\,\,\mbox{GeV}^{-2}\,.
\end{eqnarray}
Notice that the predictions of the constituent quark and vector meson
dominance models on the $\omega\to\gamma\pi (\eta)$ and $\omega\to\gamma\pi
(\eta )$ decays agree adequately with the experiment.

As is seen from Eqs. (\ref{epsilon}) and (\ref{im}), the contribution of the
$\pi\pi$ intermediate state in $Im\left (\Pi_{\omega\rho^0}(m)\right )$ and
the $g_{\omega\pi\pi}$ direct coupling constant cancel considerably in the
$g^{eff}_{\omega\pi\pi}$ effective coupling constant:
\begin{eqnarray}
\label{geff} && g^{eff}_{\omega\pi\pi}(m)=\varepsilon (m)\cdot
g_{\rho\pi\pi}+ g_{\omega\pi\pi}= -
\frac{\Pi^\prime_{\omega\rho^0}(m)\cdot  g_{\rho\pi\pi} + i\Gamma
(\rho\to\pi\pi\,,\, m)\cdot g_{\omega\pi\pi}}{m_\omega^2-m_\rho^2
+ im\left (\Gamma_\rho (m)-\Gamma_\omega (m)\right )} +
g_{\omega\pi\pi}= \nonumber\\[1pc] && = -
\frac{\Pi^\prime_{\omega\rho^0}(m)  + \left ( m_\rho^2 -
m_\omega^2 + im\Gamma_\omega (m)\right ) \cdot\left
(g_{\omega\pi\pi}/g_{\rho\pi\pi}\right )}{m_\omega^2 - m_\rho^2 +
im\left (\Gamma_\rho (m)-\Gamma_\omega (m)\right )}\cdot
g_{\rho\pi\pi}=\nonumber\\[1pc] && = -
\frac{\Pi^\prime_{\omega\rho^0}(m) \mp 1.87\cdot 10^{-4}\,
\mbox{GeV}^{-2} \pm i6.6\cdot 10^{-5}\,\mbox{GeV}^{-2}}{m_\omega^2
- m_\rho^2 + im\left (\Gamma_\rho (m)-\Gamma_\omega (m)\right
)}\cdot g_{\rho\pi\pi}\,,
\end{eqnarray}
where $\Pi^\prime_{\omega\rho^0}(m)$ is the amplitude of the $\rho^0-\omega$
transition without the contribution of the $\pi\pi$ intermediate state in
in the imaginary part, the numerical values are calculated at $m=m_\omega$.

The branching ratio of the $\omega\to\pi\pi$ decays
\begin{eqnarray}
\label{b}
&& B\left (\omega\to\pi\pi\right )=\frac{\Gamma \left (\rho\to\pi\pi\,,\,
m_\omega\right )}{\Gamma_\omega (m_\omega)}\cdot\left
|\varepsilon (m_\omega) + g_{\omega\pi\pi}/g_{\rho\pi\pi}\right |^2
\end{eqnarray}

It follows from Eqs. (\ref{gammapigammaeta}) and (\ref{geff}) that imaginary
part of the numerator in Eq. (\ref{geff}) is dominated by the $\gamma\pi$
intermediate state to within 35\%. This imaginary part gives
$B\left (\omega\to\pi\pi\right )\simeq 5\cdot 10^{-5}$ instead of the
experimental value \cite{pdg-98}
\begin{eqnarray}
\label{bexp}
B(\omega\to\pi^+\pi^-)=0.0221 \pm 0.003\,.
\end{eqnarray}
So, one can get the modulus of the real part of the numerator in Eq.
(\ref{geff}) which is clearly  dominated by
$Re\left (\Pi_{\omega\rho^0}(m)\right )$.
Besides, the interference pattern of the $\rho^0$ and
$\omega$ mesons in the $e^+e^-\to\pi^+\pi^-$ reaction and in the
$\pi^+\pi^-$ photoproduction on nuclei shows
\cite{goldhaber-69,gourdin-69,renard-70,achasov-78,achasov-92}
that the real part of the numerator in Eq. (\ref{geff}) is positive. So,
from Eqs. (\ref{epsilon}), (\ref{geff}), (\ref{b}) and (\ref{bexp}) one
obtains
\begin{eqnarray}
\label{re}
&&  Re\left (\Pi_{\rho^0\omega}(m_\omega )\right )=
( 3.80\pm 0.27 )\cdot 10^{-3}\,\mbox{GeV}^2
\end{eqnarray}
and
\begin{eqnarray}
\label{epsiloneff}
&& \varepsilon (m_\omega)+ g_{\omega\pi\pi}/g_{\rho\pi\pi}=
( 3.41\pm 0.24 )\cdot 10^{-2}\exp\left \{i\left ( 102\pm 1\right )^\circ\right \}\,.
\end{eqnarray}

The data \cite{markiii-88,dm2-90} were fitted with the function
\begin{eqnarray}
\label{fit}
N(m) = L(m) + \left |\left (N_\rho\right )^{\frac{1}{2}}F_\rho^{BW}(m) +
\left (N_\omega\right )^{\frac{1}{2}}F_\omega^{BW}(m)\exp\{i\phi\}
\right |^2\,,
\end{eqnarray}
where $F_\rho^{BW}(m)$ and $F_\omega^{BW}(m)$ are the appropriate
Breit-Wigner terms \cite{markiii-88} and $L(m)$ is a polynomial background
term.

The results are
\begin{eqnarray}
\label{experiment}
&& \phi = ( 46 \pm 15)^\circ\,,\quad N_\omega(m_\omega)/N_\rho =
8.86\pm 1.83\ \ \mbox{\cite{markiii-88}}\,,\nonumber\\
&& \phi = - 0.08\pm 0.17=(-4.58\pm 9.74)^\circ\,,\quad
N_\omega(m_\omega)/N_\rho =7.37\pm 1.72\ \ \mbox{\cite{dm2-90}}\,.
\end{eqnarray}

From Eqs. (\ref{spectrum1}), (\ref{b}), and (\ref{fit}) follows
\begin{eqnarray}
\label{Nrho}
&& N_\rho = N_\rho (m_\rho)\left |1 - \varepsilon (m_\rho)\left [
N_\omega(m_\rho)/N_\rho(m_\rho)\right ]^{\frac{1}{2}}
\exp\{i\left(\delta_\omega - \delta_\rho\right )\}\right |^2\,,\\
\label{Nomega}
&& N_\omega = B(\omega\to\pi\pi)N_\omega(m_\omega)\,,\\
\label{phi}
&& \phi = \delta_\omega - \delta_\rho +
arg\left [\varepsilon (m_\omega)+ g_{\omega\pi\pi}/g_{\rho\pi\pi}
\right ] - \nonumber \\
&& - arg\left\{1 - \varepsilon (m_\rho)\left
[N_\omega(m_\rho)/N_\rho(m_\rho)\right ]^{\frac{1}{2}}
\exp\{i\left(\delta_\omega - \delta_\rho\right )\} \right\}\simeq \nonumber\\
&& \simeq \delta_\omega - \delta_\rho + arg\left [\varepsilon (m_\omega)+
g_{\omega\pi\pi}/g_{\rho\pi\pi}\right ] - arg\left\{1 -
\left |\varepsilon (m_\omega)\right |\left [N_\omega(m_\omega)/N_\rho
\right ]^{\frac{1}{2}}\exp\{i\phi\} \right\}\,.
\end{eqnarray}

From Eqs. (\ref{epsiloneff}), (\ref{experiment}) and (\ref{phi}) we obtain
\begin{eqnarray}
\label{markiii}
&& \delta_\rho - \delta_\omega = \delta_\gamma = ( 60\pm 15 )^\circ\ \
\mbox{\cite{markiii-88}}\,,\\
\label{dm2}
&& \delta_\rho - \delta_\omega = \delta_\gamma = ( 106\pm 10 )^\circ\ \
\mbox{\cite{dm2-90}}\,.
\end{eqnarray}

A large ( nearly $90^\circ$ ) $\delta_\gamma$ was obtained in Ref.
\cite{castro-95,suzuki-98}. So, both the MARK III Collaboration
\cite{markiii-88} and the DM2 Collaboration \cite{dm2-90}, see Eqs.
(\ref{markiii}) and (\ref{dm2}), provide support for this view.

The DM2 Collaboration used statistics only half as high as the MARK III
Collaboration, but, in contrast to the MARK III Collaboration, which fitted
$N_\omega$ as a free parameter, the DM2 Collaboration calculated it from
the branching ratio of $J/\psi\to\omega\eta$ using Eq. (\ref{Nomega}).

In summary we would like to emphasize that it would be beneficial to study
this fundamental problem once again with BES in Beijing.

We thank G.F. Xu very much for discussions.
The present work was supported in part by the grant INTAS-RFBR IR-97-232.

\end{document}